\documentclass[twocolumn,groupedaddress,amsmath,amssymb,prl,aps]{revtex4}

\usepackage{graphicx}
\usepackage{dcolumn}
\usepackage{bm}
\usepackage{soul} 
\usepackage{xcolor} 
\usepackage{amsmath} 

\begin{document}

\title{Creating currents of electric bubbles}

\author{Jorge \'I\~niguez-Gonz\'alez$^{1,2}$ and Hugo Aramberri$^{1}$}

\affiliation{
  \mbox{$^{1}$Luxembourg Institute of Science and Technology (LIST),}
  \mbox{Avenue des Hauts-Fourneaux 5, L-4362 Esch/Alzette,
    Luxembourg}\\
 \mbox{$^{2}$Department of Physics and Materials Science, University
   of Luxembourg, Rue du Brill 41, L-4422 Belvaux, Luxembourg}}

\maketitle   

\textbf{The experimental demonstration of electric skyrmion bubbles
  and the recent prediction of their Brownian motion have brought
  topological ferroelectrics close to their magnetic
  counterparts. Electric bubbles (e-bubbles) could potentially be
  leveraged in applications for which magnetic skyrmions have been
  proposed (e.g., neuromorphic computing). Yet, we still lack a
  strategy to create currents of e-bubbles. Here, using predictive
  atomistic simulations, we illustrate two approaches to induce
  e-bubble currents by application of suitable electric fields, static
  or dynamic. We focus on regimes where e-bubbles display spontaneous
  diffusion, which allows us to generate a current by simply biasing
  their Brownian motion. Our calculations indicate that e-bubble
  velocities over 25~m/s can be achieved at room temperature,
  suggesting that these electric quasiparticles could rival the speeds
  of magnetic skyrmions upon further optimization.}


Recent studies have addressed the possibility of stabilizing and
manipulating electric bubbles
(e-bubbles)~\cite{lichtensteiger14,zhang17,bakaul21,prokhorenko24,bastogne24},
evidenced their nontrivial topologies~\cite{das19,goncalves19,han22},
and even predicted their behavior as Brownian
quasiparticles~\cite{aramberri24}. Perovskites are the best studied
materials, with ferroelectric/dielectric PbTiO$_{3}$/SrTiO$_{3}$
superlattices (PTO/STO) emerging as model systems~\cite{junquera23}.

The prediction that e-bubbles can display spontaneous stochastic
diffusion resonates with novel computing concepts that leverage
thermal noise and the associated
randomness~\cite{peper13,pinna18,grollier20,kaiser21}. Ultralow-power
devices based on Brownian magnetic skyrmions have been proposed, and
functionalities ranging from non-linearly separable XOR operations
\cite{raab22,cortes95} to pattern recognition \cite{yokouchi22}
demonstrated. These schemes fall within the category of
``unconventional computing'', a promising path toward sustainable
artificial intelligence~\cite{finocchio24}. In this context, e-bubbles
controllable by electric fields -- as opposed to the electric currents
needed to act on magnetic skyrmions -- may enable more efficient
devices.

To fulfill this promise, though, we must improve our understanding of
e-bubbles and, in particular, address their mobility. Since 2010 we
know that small electric currents can be used to move magnetic
skyrmions~\cite{jonietz10,nagaosa13}. Can something equivalent be done
with e-bubbles? Here we give a postive answer to this question,
presenting simulation evidence for e-bubble currents driven by
suitable electric fields.


We work with PTO/STO superlattices where e-bubbles have been
observed~\cite{das19,aramberri24}, using the atomistic
``second-principles'' simulation techniques \cite{wojdel13,zubko16}
that have successfully predicted the main properties of these
materials~\cite{shafer18,goncalves19,das19,graf22}. (See Methods for
details.) A compressive epitaxial strain in the $xy$ plane, e.g. as
imposed by a STO substrate, yields an easy polar axis along the
stacking direction $z$. Further, the dielectric STO layers impose
open-circuit-like electric boundary conditions on the PTO layers,
which develop ferroelectric stripe domains with $P_{z} = 0$
overall. If an electric field is applied along $z$, the stripes break
and e-bubbles form~\cite{aramberri24}. Typically, the bubbles span the
whole thickness of the PTO layer and have an approximately circular
$xy$ section a few nanometers in diameter.

In ultrathin PTO layers (below 10 unit cells) the bubbles have a small
surface, which suggests that thermal fluctuations may result in a net
drift of the whole object. Indeed, simulations predict that -- within
a range of about 100~K below the Curie point -- e-bubbles diffuse
spontaneously~\cite{aramberri24,gomez-ortiz24}, an effect consistent
with XRD data~\cite{zubko16}. The bubbles are predicted to remain
stable even when diffusing, behaving as long-lived Brownian
particles~\cite{aramberri24}.

This leads to a self-evident notion: if we create an asymmetry in the
$xy$ plane, the Brownian diffusion must yield a net bubble
current. Indeed, following similar ideas, a temperature gradient was
used to move magnetic skyrmions~\cite{raimondo22}. Here we use
electric fields to drive the effect, as they are an experimentally
convenient choice.


Let us first show how to induce a current of e-bubbles by applying a
static position-dependent electric field. We apply a $z$-oriented
homogeneous field ${\cal E}^{(0)}_{z}$ to create bubbles in the PTO
layers. Assume ${\cal E}^{(0)}_{z} >0$, so the bubbles correspond to
regions of downward polarization within a matrix polarized upward. Now
consider a linear spatial dependece of the total field,
\begin{equation}
    {\cal E}_{{\rm tot},z}(x) = {\cal E}_{z}^{(0)} + {\cal
      E}_{z}^{(1)}\frac{2x-L}{2L} \; ,
  \label{eq:field}
\end{equation}
where $L$ is a characteristic length of our simulated system, ${\cal
  E}_{z}^{(1)}/L$ quantifies a field gradient, and we assume ${\cal
  E}_{z}^{(1)}>0$. If we approximate the bubble as a point dipole
$d_{b,z}<0$, its energy under this field is given by $V_b \approx -
{\cal E}_{{\rm tot},z} \, d_{b,z}$. The bubble thus experiences a
force
\begin{equation}
  f_{b,x} = - \frac{\partial V_{b}}{\partial x} = \frac{{\cal
      E}_{z}^{(1)}d_{b,z}}{L} < 0 \; ,
  \label{eq:force}
\end{equation}
which drives it toward smaller $x$ values. Hence, in principle, a
field gradient may allow us to move e-bubbles.

To test this, we run simulations of PTO/STO superlattices that we
denote 9/3 and 6/3 -- i.e., where the STO layers have a thickness of 3
elemental perovskite cells, while the PTO layers have thicknesses of 9
and 6 cells, respectively. We work in conditions where long-lived
Brownian bubbles were previously predicted~\cite{aramberri24}, namely,
with the 9/3 superlattice at around 300~K and subject to a $-2$\%
compressive epitaxial strain, and with the 6/3 system at around 200~K
and a $0$\% epitaxial strain. (The STO lattice constant is taken as
the zero of strain.) We consider simulation supercells that can be
seen as an $N\times 8 \times 1$ repetition of the elemental
superlattice unit, and impose periodic boundary conditions. While we
consider $N$ values between 8 and 32, most of our results are for $N =
8$, which proves sufficient. The applied ${\cal E}^{(0)}_{z}$ fields
are chosen to stabilize a single e-bubble in the $8\times 8\times 1$
supercell; we find this can be done for ${\cal E}^{(0)}_{z}$ around
1.0~MV~cm$^{-1}$ and 1.5~MV~cm$^{-1}$, respectively, for the 9/3 and
6/3 systems.

\begin{figure}
\includegraphics[width=0.75\columnwidth]{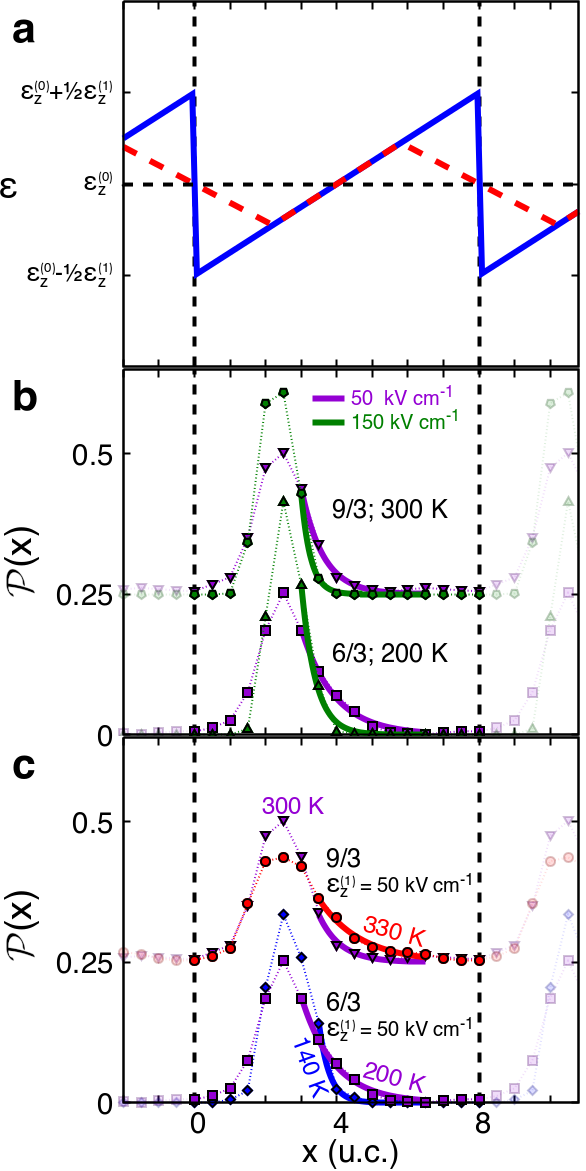}
\caption{{\bf Confined e-bubbles under the action of a
    sawtooth-modulated electric field.} In panel~{\bf a}, the blue
  solid line shows the variation with $x$ of the sawtooth-modulated
  electric field, in a periodically-repeated supercell with a length
  of $N=8$ perovskite units. The red dashed line is proportional to
  the electric potential ($\tilde{V}_{b}$) experienced by an e-bubble
  of 4~unit cells in diameter, as caused by the applied field (blue).
  Panels~{\bf b} and {\bf c} show the probability distribution for the
  e-bubble position (${\cal P}(x)$) for varying modulation amplitude
  and temperature, respectively.  In each panel, the top lines
  correspond to the 9/3 superlattice ($-2$~\% epitaxial strain, ${\cal
    E}_{z}^{(0)}= 1.0$~MV~cm$^{-1}$) while the bottom lines pertain to
  the 6/3 system (no epitaxial strain, ${\cal E}_{z}^{(0)}=
  1.5$~MV~cm$^{-1}$).  Symbols correspond to computed data, thin
  dotted lines are guides to the eye, and solid thick lines are fits
  to Eq.~(\protect\ref{eq:exp}).}
\label{fig:prob}
\end{figure}

As sketched in Figure~1{\bf a}, the use of periodic boundary
conditions forces us to apply a sawtooth-modulated field, rather than
a homogeneous gradient. This field is given by Eq.~(\ref{eq:field})
for $0<x<L$, where $L$ is the length of the simulation supercell along
$x$, and it is periodically repeated for $x<0$ and $x>L$. Thus, ${\cal
  E}^{(1)}_{z}$ quantifies the field change across one supercell
period. Typically we consider ${\cal E}^{(1)}_{z}$ values within 20\%
of the homogeneous component ${\cal E}^{(0)}_{z}$; such perturbations
do not change the bubble density in our supercell.

We run molecular dynamics (MD) simulations to study the equilibrium
state of these systems. (See Methods for details.) In particular, we
compute ${\cal P}(x)$, the probability of finding the bubble centered
at $x$. Figures~1{\bf b} and 1{\bf c} summarize our results. We find
that the bubble is largely restrained to the left half of the
simulation supercell. As expected, the confinement is stronger as
${\cal E}_{z}^{(1)}$ increases (Fig.~1{\bf b}) and at lower
temperatures (Fig.~1{\bf c}).

Interestingly, here we have a drift-diffusion problem where a Brownian
particle is subject to a time-independent potential
$\tilde{V}_{b}(x)$, as sketched in Fig.~1{\bf a}. $\tilde{V}_{b}(x)$
differs from the ideal potential $V_{b}(x) \approx - {\cal E}_{{\rm
    tot},z} \, d_{b,z}$ mentioned above because of the periodic
boundary conditions in our simulations and the fact that e-bubbles are
not point dipoles. Nevertheless, $\tilde{V}_{b}(x)$ can be expected to
have an approximately linear dependence with $x$ in the central region
of the supercell, resulting in a constant drift force
$\tilde{f}_{b,x}(x) \propto {\cal E}^{(1)}_{z}$ acting on the
bubble. As argued in Supplementary Note~1, within that constant-drift
region we expect ${\cal P}(x)$ to satisfy a simplified Smoluchowski
equation~\cite{balakrishnan-book21},
\begin{equation}
  c \frac{\partial{\cal P}(x)}{\partial x} + D \frac{\partial^{2}{\cal
  P}(x)}{\partial x^{2}} = 0 \; ,
\end{equation}
where $c$ is the drift velocity and $D$ the diffusion constant of the
Brownian bubble. This equation is solved for
\begin{equation}
  {\cal P}(x) = A \exp{\left(-\frac{c}{D}x\right)} \; ,
  \label{eq:exp}
\end{equation}
where $A$ is an integration constant. We thus expect ${\cal P}(x)$ to
approach zero exponentially fast as we move into the right half of the
supercell, the decay being controlled by $c/D$.

\begin{figure}
\includegraphics[width=0.95\columnwidth]{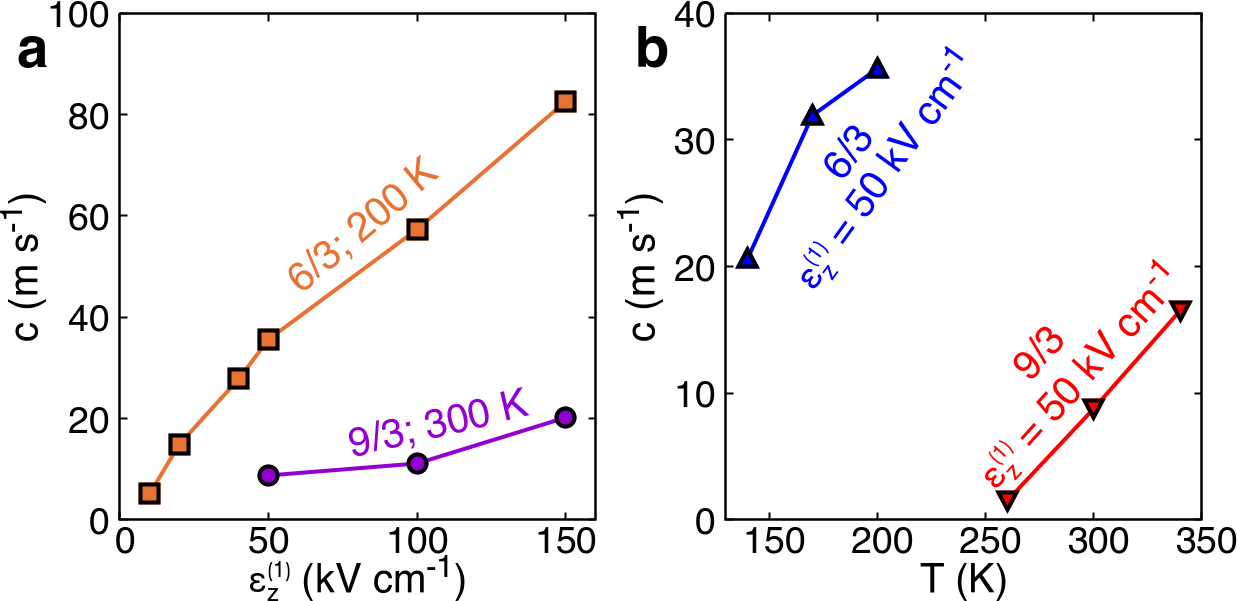}
\caption{{\bf Predicted e-bubble velocities under static bias.} The
  drift velocities obtained by fitting to Eq.~(\ref{eq:exp}) grow as
  we increase the amplitude of the sawtooth modulation ({\bf a}) or
  the temperature ({\bf b}). The results correspond, essentially, to
  the cases displayed in Fig.~\protect\ref{fig:prob}.}
\label{fig:velocities-gradient}
\end{figure}

As shown in Fig.~1, we can fit well the tails of the computed ${\cal
  P}(x)$ probability densities using Eq.~(\ref{eq:exp}), and thus
obtain the corresponding $c/D$ ratios. Using $D$ obtained from MD
simulations with ${\cal E}^{(1)}_{z} = 0$, we can calculate the
velocities $c$. Our results, summarized in Figure~2, show an
approximate proportionality between $c$ and ${\cal E}^{(1)}_{z}$ (as
expected; see Supplementary Note~1) and a significant $T$
dependence. Remarkably, we predict e-bubble velocities over 20~m/s.


This indirect approach to estimate the bubble velocity $c$ may seem
tenuous. Can we actually observe e-bubbles moving at 20~m/s in our
simulations? To answer this, we now consider a time-dependent electric
field of the form
\begin{equation}
  {\cal E}_{{\rm tot},z}(x;t) = {\cal E}_{z}^{(0)} + {\cal
    E}^{(1)}_{z}\, {\rm sin}\left( \frac{2\pi}{L}x - \frac{2\pi}{\tau}
  t \right) \; .
  \label{eq:wave}
\end{equation}
Here, $\tau$ is the time it takes the sinusoidal perturbation to
complete a full oscillation; hence, $v_{\rm W} =L/\tau$ is the
field-wave velocity. Noting that we work with $L \approx 8 \times
3.9$~\AA, a field wave propagating at 20~m/s corresponds to $\tau
\approx 150$~ps, a time scale within the scope of our
second-principles calculations~\cite{aramberri24}.

\begin{figure}
\includegraphics[width=0.95\columnwidth]{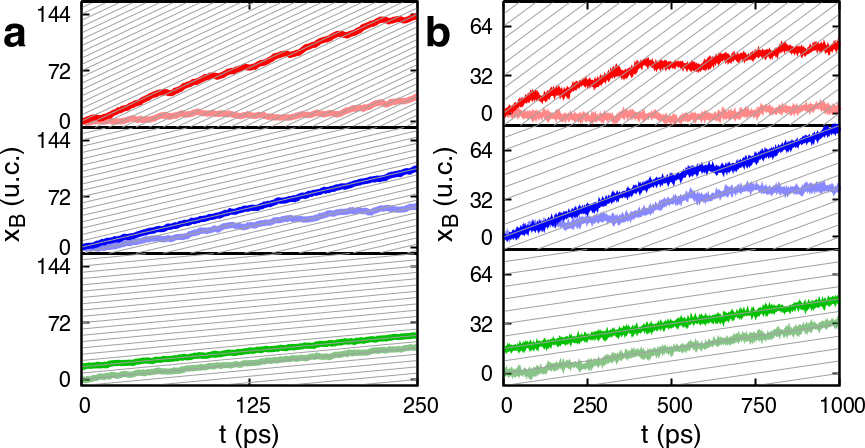}
\caption{{\bf Simulated e-bubble motion driven by a field wave.}
  Bubble position ($x_{\rm B}$) as a function of time. Representative
  results for 6/3 and 9/3 superlattices are shown in panels~{\bf a}
  and {\bf b}, respectively. In every subpanel, the data shown with
  brighter colors correspond to a wave amplitude ${\cal E}^{(1)}_{z} =
  100$~kV~cm$^{-1}$, while dim colors are used for ${\cal E}^{(1)}_{z}
  = 50$~kV~cm$^{-1}$. The background lines track the motion of the
  minima of the sinusoidal field modulation; such minima correspond to
  regions where the e-bubble has a lower potential energy. For slow
  waves (bottom subpanels, green lines) the e-bubbles follow the field
  wave, even for modulations of small amplitude. As the waves get
  faster (medium and top subpanels), the e-bubbles begin to get off
  track sometimes, to the point that they may become nearly immobile
  (see cases in dim red, top subpanels).}
\label{fig:current}
\end{figure}

We thus consider waves with amplitudes ${\cal E}^{(1)}_{z}$ within
20\% of the homogeneous component ${\cal E}^{(0)}_{z}$, and $\tau$'s
between 10~ps and 1000~ps. The results in Figure~3 illustrate the
three different situations we find: bubbles following comfortably a
slow wave (green), bubbles having trouble to follow a faster wave and
getting off track occasionally (blue), and bubbles incapable of
keeping up with very fast waves (red). These regimes can be best
appreciated in Supplementary Movies 1 to 3, respectively.

\begin{figure}
\includegraphics[width=0.95\columnwidth]{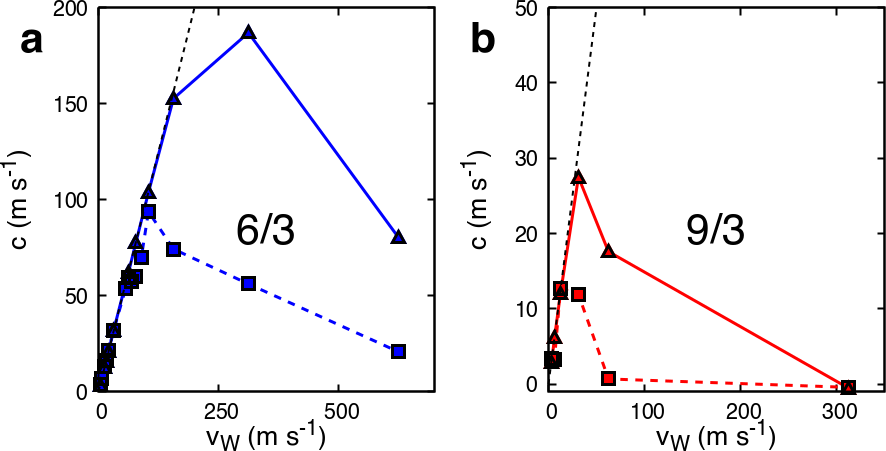}
\caption{{\bf Predicted e-bubble velocities under a field wave.} We
  show results for our model 6/3 ({\bf a}) and 9/3 ({\bf b})
  superlattices, using solid and dashed lines for wave modulations
  with ${\cal E}^{(1)}_{z} = 100$~kV~cm$^{-1}$ and ${\cal E}^{(1)}_{z}
  = 50$~kV~cm$^{-1}$, respectively. The black dashed line corresponds
  to a perfect tracking, where $c = v_{\rm W}$. The results yield a
  clear qualitative picture; note, though, that obtaining good
  statistics in the regime where $c\lesssim v_{\rm W}$ would require
  prohibitively long simulations, as in such cases the bubbles'
  slip-offs are rare events.}
\label{fig:current-v}
\end{figure}

Figure~4 shows the average e-bubble velocity, $c$, against the wave
velocity $v_{\rm W}$. In the representative case of the 6/3
superlattice, we find $c \approx v_{\rm W}$ at small wave
velocities. This perfect tracking regime extends up to about 50~m/s
for a relatively small perturbation with ${\cal E}_{z}^{(1)} =
50$~kV~cm$^{-1}$, and up to about 150~m/s for ${\cal E}_{z}^{(1)} =
100$~kV~cm$^{-1}$. Then, $c$ always reaches a maximum beyond which the
bubbles slow down considerably. We obtain maximum bubble speeds of
about 180~m/s and 30~m/s, respectively, for the 6/3 superlattice at
200~K and the 9/3 superlattice at 300~K. We thus ratify the prediction
of fast e-bubble currents. Remarkably, the bubbles remain stable
throughout the simulations, even when they are dazed by fast field
waves, which confirms their resilience as long-lived quasiparticles.


It thus seems that field waves allow us to drag e-bubbles faster than
static gradients do. The reason is simple: for the same ${\cal
  E}_{z}^{(1)}$, the field waves exert on the bubbles a force (given
essentially by the $x$-derivative of Eq.~(\ref{eq:wave})) that can be
up to $2\pi$ times larger than the maximum force caused by the field
gradient (Eq.~(\ref{eq:force})).

When are the waves too fast for the e-bubbles to follow? Recall that
e-bubbles move by switching dipoles at their boundary; hence, their
velocity is ultimately limited by the speed of such a local
polarization switching. A bubble velocity of 100~m/s implies that
boundary cells switch in about 4~ps. This may seem surprisingly fast;
yet, it is consistent with atomistic studies of ferroelectric
switching in PTO~\cite{shin07}. Moreover, the authors of
Ref.~\onlinecite{shin07} found that field-driven domain wall motion in
PTO proceeds through the formation of a critical nucleus of about
$3\times 3$ unit cells. Noting that the switching of interest here
involves only a few boundary cells, it is conceivable it will
typically comprise a single nucleation event, thus being ultrafast.

Hence, while intriguing aspects of the simulated bubble currents
remain for future study (e.g., concerning the $T$-dependence of $c$),
our basic results seem physically sound and yield a clear picture. In
Supplementary Note~2 and Supplementary Figures S1 to S3 we present
additional results further supporting our main conclusions.

The cases simulated here are admittedly challenging (we are restricted
by the excessive computational cost of treating bigger systems); yet,
we believe they fall within the range of what is experimentally
conceivable. For example, it may be possible to create suitable static
field gradients by using wedge-shaped electrodes, taking advantage of
advances in nanofabrication~\cite{berenschot22}. Also, progress on
surface acoustic waves suggests that field waves propagating at
hundreds of m/s with wavelenghts of tens of nm may soon be within
reach~\cite{delsing19}. Pinning, though, will constitute an
unavoidable difference between our defect-free simulations and
experiments. Pinning will reduce the number of mobile bubbles and slow
down the ones that can move. Nevertheless, we see no reason to believe
that all e-bubbles will be clamped in high-quality samples -- note
e.g. the experimental results of Ref.~\cite{zubko16}, which suggest
stochastic dynamics of ferroelectric domains. Hence, we think that
ultimately pinning will not be an unsurmountable problem.

We conclude by noting that, typically, magnetic skyrmions can be
accelerated to about 100~m/s, and only recently \cite{pham24} have
there been reports of higher speeds (900~m/s). Notwithstanding the
differences between our theoretical e-bubbles and the magnetic
skyrmions actually measured, it is remarkable that our predictions --
where no velocity optimization was attempted -- yield results on par
with record-setting magnetic systems. Hence, our calculations suggest
that e-bubbles may become a quasiparticle of choice in applications
where magnetic skyrmions are being considered, e.g. for ultralow-power
neuromorphic computing. We hope these results will attract the
attention of physicists and engineers alike, as they may herald an
exciting era of research -- fundamental and applied -- focused on
electric-bubble currents.


\textbf{Methods}

Second-principles simulations are performed using the SCALE-UP package
\cite{wojdel13,garciafernandez16,scaleup} and the same approach as in
previous studies of PTO/STO
superlattices~\cite{zubko16,shafer18,das19}. The superlattice models
are based on potentials for the pure bulk compounds -- fitted to
first-principles results \cite{wojdel13} -- and adjusted for the
superlattices~\cite{zubko16}. The e-bubble simulations and analysis
(e.g., definition of e-bubble centers, quantification of their
trajectories) follow the methodology described in
Ref.~\onlinecite{aramberri24}. The only noteworthy differences pertain
to the MD simulations with electric-field waves. In such cases, we
first prepare a thermalized state (atomic positions and velocities) by
running an isokinetic MD simulation of the material under the action
of a static wave-modulated electric field (i.e., as obtained from
Eq.~(\ref{eq:wave}) for $t=0$). Then, we turn on the motion of the
electric field wave and simulate the system dynamics by simply
following Newton's equations of motion. Note, though, that here we
have a time-dependent potential, so the total energy of the system is
not conserved. To keep the temperature at its desired value, we
thermostat the system by applying a suitable velocity rescaling every
50~ps. We explicitly check that such a rescaling has no significant
effect on the e-bubble diffusion. All our MD simulations with
sawtooth-modulated fields run for at least 3~ns, which we find is
enough to get reliable results for ${\cal P}(x)$. All our MD
simulations with field waves run for at least 1~ns.

\vspace{5mm}{\bf Acknowledgements}

Work funded by the Luxembourg National Research Fund (FNR), mainly
through grant C21/MS/15799044/FERRODYNAMICS and also through grant
C23/MS/17909853/BUBBLACED. We are thankful for inspiring discussions
with the members of the TOPOCOM Marie Skłodowska-Curie Doctoral
Network, particularly D.R. Rodrigues (Bari).

\end{document}